\documentclass[prl,superscriptaddress,showpacs,aps,twocolumn]{revtex4-1}
\usepackage{amsmath}
\usepackage{graphicx}
\usepackage{dcolumn}% Align table columns on decimal point
\usepackage{bm}% bold math
\usepackage{epsfig}
\usepackage{color}

\newcommand{\Rvec}{{\bf R}}

\newcommand{\qvec}{{\bf q}}
\newcommand{\rvec}{{\bf r}}
\newcommand{\xvec}{{\bf x}}
\newcommand{\Xvec}{{\bf X}}

\newcommand{\Qvec}{{\bf Q}}
\newcommand{\beq}{\begin{equation}}
\newcommand{\eeq}{\end{equation}}
\newcommand{\bea}{\begin{eqnarray}}
\newcommand{\angm}[1]{\AA$^{-#1}$}

\begin{document}
\title{Nonlinear Network description for many-body quantum systems in continuous space}

\author{Michele Ruggeri}
\affiliation{Max Planck Institute for Solid State Research, Heisenbergstr. 1, 70569 Stuttgart, Germany}
\author{Saverio Moroni}
\affiliation{DEMOCRITOS National Simulation Center, Istituto Officina dei Materiali del CNR and SISSA, 
Via Bonomea 265, I-34136 Trieste, Italy}
\author{Markus Holzmann}
%\affiliation{LPTMC, UMR 7600 of CNRS,  Universit{\'e} Pierre et Marie Curie, Paris, France}
\affiliation{LPMMC, UMR 5493 of CNRS, Universit{\'e} Grenoble Alpes, F-38100 Grenoble
France}
\affiliation{Institut Laue Langevin, BP 156, F-38042 Grenoble Cedex 9, France}

\date{\today}
\begin{abstract}
We show that the recently introduced iterative backflow 
renormalization can be interpreted as a general neural network in continuum 
space with non-linear functions in the hidden units.
We use this wave function within Variational Monte Carlo for liquid 
$^4$He in two and three dimensions, where we typically find a tenfold
increase in accuracy over currently used wave functions.
Furthermore, subsequent stages of the iteration
procedure define a set of increasingly good wave functions,
each with its own variational energy and variance of the local energy:
extrapolation of these energies to zero variance gives values 
in close agreement with the exact values. For two dimensional $^4$He, we 
also show that the iterative backflow wave function can describe 
both the liquid and the solid phase with the same functional form 
--a feature shared with the Shadow Wave Function, but 
now joined by much higher accuracy.
We also achieve significant progress for liquid $^3$He in three dimensions,
improving previous variational and fixed-node energies
for this very challenging fermionic system.
\end{abstract}
\pacs{PACS: }
\maketitle

Explicit forms of many-body ground state wave functions have played an important role in the qualitative and quantitative understanding of many-body quantum systems. Whereas pairing functions based on Bogoliubov's theory \cite{Bogoliubov} have provided a good description of superfluidity and superconductivity of dilute gases, a full pair-product (Jastrow) wave function
is usually the starting point for a microscopic description of liquid helium,
the prototype of a strongly interacting, correlated quantum system.
Starting from the first variational Monte Carlo (VMC) calculations of McMillan \cite{McMillan}, 
liquid and solid helium -- bosonic $^4$He as well as fermionic $^3$He --
have triggered and challenged microscopic simulations to describe many-body quantum systems
in two or three dimensional continuous space.

For systems described on a lattice, approaches based on
matrix product and tensor network states \cite{MPISFannes,MPSWhite,MPS,MPS2,TNN} have provided essentially exact description of many generic
low dimensional systems. Very recently, neural network states have been shown to lead to excellent results in one and two dimensional lattice models \cite{CT,NNHubbardBose1,NNHubbardBose2,NNHubbardFermi}, a very promising approach for lattice systems
in two and three dimensions. However, generalization of these states to continuous systems \cite{cMPS} in two
and three dimensions is difficult or still lacking.

In this work we elaborate on a recently introduced \cite{2d} class of wave functions for quantum many-body systems in 
continuous space that include sets of auxiliary coordinates obtained
with iterated backflow transformations. The wave function is viewed as a neural network
where the hidden units of layer $M$ are obtained iteratively as a function of the coordinates in layer
$M-1$, with layer $M=0$ corresponding to the physical particles. In contrast to neural networks on a lattice, all the functions involved here are in general non-linear. The network parameters describing the various functions are optimized within VMC simulations. 

We apply our description to liquid/solid $^4$He and liquid $^3$He, where we obtain a systematic lowering of the energy as we increase the number of layers in the wave functions.
For the bosonic systems we benchmark the quality of this explicit wave function with
exact results obtained by stochastic projection Monte Carlo methods. We further show that
our wave function is able to describe equally well the fluid and the solid phase with the same functional form, symmetric and translationally invariant,
qualitatively similar to the so-called {\em shadow wave function} (SWF) approach\cite{vitiello} but with over one order of magnitude gain in accuracy.

Since the effective interaction between
two helium atoms, $v(r)$, is quantitatively well known, a large quantity of computations exists which
can be rather directly compared to experiments.
During the years several types of wave functions have been used to simulate $^4$He.
In the first VMC simulations \cite{McMillan}, the wave function took into account just two-body interparticle correlations; 
these wave functions were then generalized to include three-body and higher correlations, $\Psi_T(\Rvec) \propto \exp[-U(\Rvec)]$, where $U(\Rvec)$ denotes a general, symmetric correlation function, and $\Rvec \equiv (\rvec_1, \rvec_2,\dots \rvec_N)$ denotes the coordinate vector of the particles \cite{slk,manyBodyCorr}.

\begin{figure}
\includegraphics[width=\columnwidth]{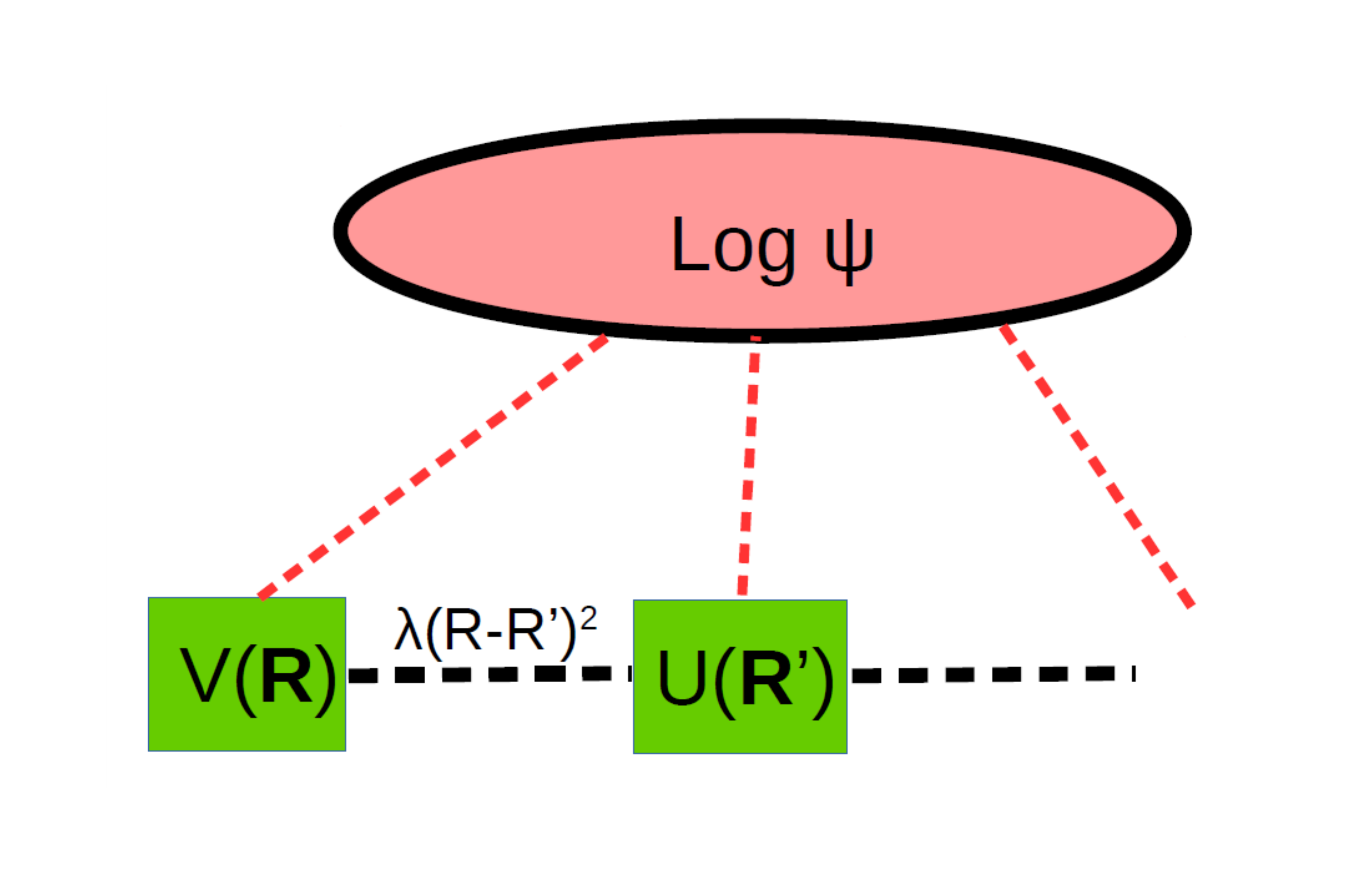}
\caption{Schematic representation of a SWF as a non-linear network. The input layer is formed by the coordinates of the wave function, $\Rvec$, and we have to integrate over the coordinates in the hidden layer, $\Rvec'$. Input and hidden layer coordinates are connected via a gaussian, whereas the coordinates inside each layer are connected by the many-body correlation potentials, $V(\Rvec)$ and $U(\Rvec')$. Including several hidden layers correspond to the application over several projection steps.}
\label{SWF}
\end{figure}

Exact results for bosonic $^4$He can be obtained improving stochatically the wave function
with Projector Monte Carlo techniques such as Diffusion Monte Carlo (DMC) \cite{mitas} or Variational Path Integral methods\cite{RMP,pigs,rqmc}. 
Starting from any trial wave function, $\Psi_T(\Rvec)$, its propagation in imaginary time, $\tau$, can be written as
\begin{equation}
\Psi_\tau(\Rvec) \propto \int d \Rvec' G(\Rvec, \Rvec';\tau) \Psi_T(\Rvec')
\label{psitau}
\end{equation}
For small $\tau$, the functional form of $G$ is given by 
\begin{equation}
G(\Rvec,\Rvec'; \tau) \propto \exp \left[ - \lambda (\Rvec-\Rvec')^2 -V(\Rvec) \right]
\label{gtau}
\end{equation}
with $\lambda=m/2\hbar^2 \tau$ and $V(\Rvec)$ is given by the interparticle potential, $V(\Rvec)=\tau \sum_{i<j} v(r_{ij})$. Large 
projection times can be reached by iterative application of the short time propagator; the integrals can then be sampled numerically via projection Monte Carlo methods. 

Alternatively, we can consider Eqs~(\ref{psitau}) and (\ref{gtau}) as an improved variational ansatz for our ground state, the SWF \cite{vitiello}, and minimize the energy with respect to variations in $\lambda$, $V$, and $U$. 
In contrast to the explicit trial wave functions, $\Psi_T$,  used in previous VMC calculations,
the SWF is able to describe the melting from solid to liquid $^4$He without modification of the structure.

Both shadow and projector Monte Carlo methods explicitly depend on auxiliary (or hidden) variables, $\Rvec'$. The resulting wave function thus forms a network where the hidden variables are connected to the input layer, $\Rvec$. However, in contrast to many neural network systems on a lattice,
the variables inside of each layer are connected to each other via the many-body potentials, $V(\cdot)$ and $U(\cdot)$. Figure \ref{SWF} shows a schematic representation of the SWF network.

Within SWF and projection Monte Carlo methods, the integration over the variables in the hidden layers is done stochastically. 
This will in general lead to a sign (phase) problem whenever $G$ or $\psi_T$ carries a sign (phase) as for fermionic or time-dependent problems \cite{shadowsHe3,shadowsHe3-2,shadowsign}. 
Analytical integration over the hidden layer then becomes extremely important, since the evaluation
of the resulting explicit form may be possible within a standard VMC approach without facing a sign problem.

In our case, the integration over the hidden variables cannot be done analytically.
However, we can approximately perform the integrations in Eq.~(\ref{psitau}) expanding $ U(\Rvec')$ around some positions $\Qvec$, which will be fixed later. For large $\lambda$, we can truncate the expansion after the linear term 
\begin{eqnarray}
\Psi_\tau(\Rvec)& \approx& \int d\Rvec'
\exp \left[
- \lambda \left(\Rvec'-\Rvec + \nabla U /2\lambda \right)^2
- V(\Rvec)\right] \nonumber \\
&\times & \exp \left[-U- (\Rvec-\Qvec) \cdot \nabla U + (\nabla U )^2/4\lambda
\right]
\label{Psitau1}
\end{eqnarray} 
where $U$ and $\nabla U$ are evaluated at $\Qvec$.
The gaussian integrals are centered around 
\begin{equation}
\Qvec = \Rvec -  \nabla U(\Qvec) /2\lambda
\label{Qvec}
\end{equation}
which gives an implicit equation to determine $\Qvec$. 

Performing the gaussian integration, we get
\begin{equation}
\Psi_\tau(\Rvec) \sim  
 \exp \left[-V(\Rvec) -U(\Qvec)- [\nabla U(\Qvec) ]^2/4\lambda
\right]
\label{Psitau2}
\end{equation}
The resulting wave function can then be  put into the form 
\begin{equation}
\Psi_\tau(\Rvec) = \Phi^{(0)}(\Rvec) \cdot \Phi^{(1)}\left(\Qvec\right) 
\label{wf1}
\end{equation}
where $\Phi^{(n)}(\cdot)=\exp[-U^{(n)}(\cdot)]$ is a correlated wave function containing generalized many-body Jastrow potentials, $U^{(n)}(\cdot)$. Although our derivation suggests explicit expressions for $U^{(n)}$ and $\Qvec$ in terms of $V$, $U$, and $\lambda$, we rather retain only the functional form, and we 
simplify Eq.~(\ref{Qvec}) by replacing $\Qvec$ with $\Rvec$ in the r.h.s. The corresponding parameters are then optimized, such that the wave function, Eq.~(\ref{wf1}), minimizes some target function, usually taken as the energy or the variance of the local energy.

In projection Monte Carlo algorithms, the exact ground state is obtained 
by iterative applications of the propagator, $G$.
Similarly, we want to improve our wave function by approximately applying 
the propagator to $\Psi_\tau$.
If we again identify $\Rvec$ with $\Qvec$ in Eq.~(\ref{wf1}), $\Psi_\tau$ 
is of similar form as the original trial wave function, $\Psi_T$,
namely the exponential of generalized Jastrow potentials, and we can
apply the derivation outlined above, Eqs.~(\ref{Psitau1}--\ref{Psitau2}), to obtain
$\Psi_{2\tau}$.
We end up with a rather simple, iterative structure
\begin{equation}
\Psi^{(M)}(\Rvec) = \prod_{n=0}^M \Phi^{(n)}\left(\Qvec^{(n)}\right) 
\label{wf}
\end{equation}
where $M$ is the number of iterative backflow transformation. 
At each level, $n$, new backflow coordinates are introduced
\begin{equation}
\Qvec^{(n)}= \Qvec^{(n-1)} + \nabla \widetilde{U}^{(n-1)}(\Qvec^{(n-1)})
\label{Qvecn}
\end{equation} 
These generalized backflow coordinates are iteratively built from the backflow and Jastrow potentials of the previous level, $\Qvec^{(n-1)}$ and  
$\widetilde{U}^{(n-1)}$, respectively, starting from
$\Qvec^{(0)} \equiv \Rvec$. The notation $\widetilde{U}^{(n)}$ indicates that we may use the
same functional form in all the generalized Jastrow 
factors $U^{(n)}\equiv -\log \Phi^{(n)}$. 
In practice, we have used the simplest possible two- and three-body forms for our explicit calculations below.

\begin{figure}
\includegraphics[width=\columnwidth]{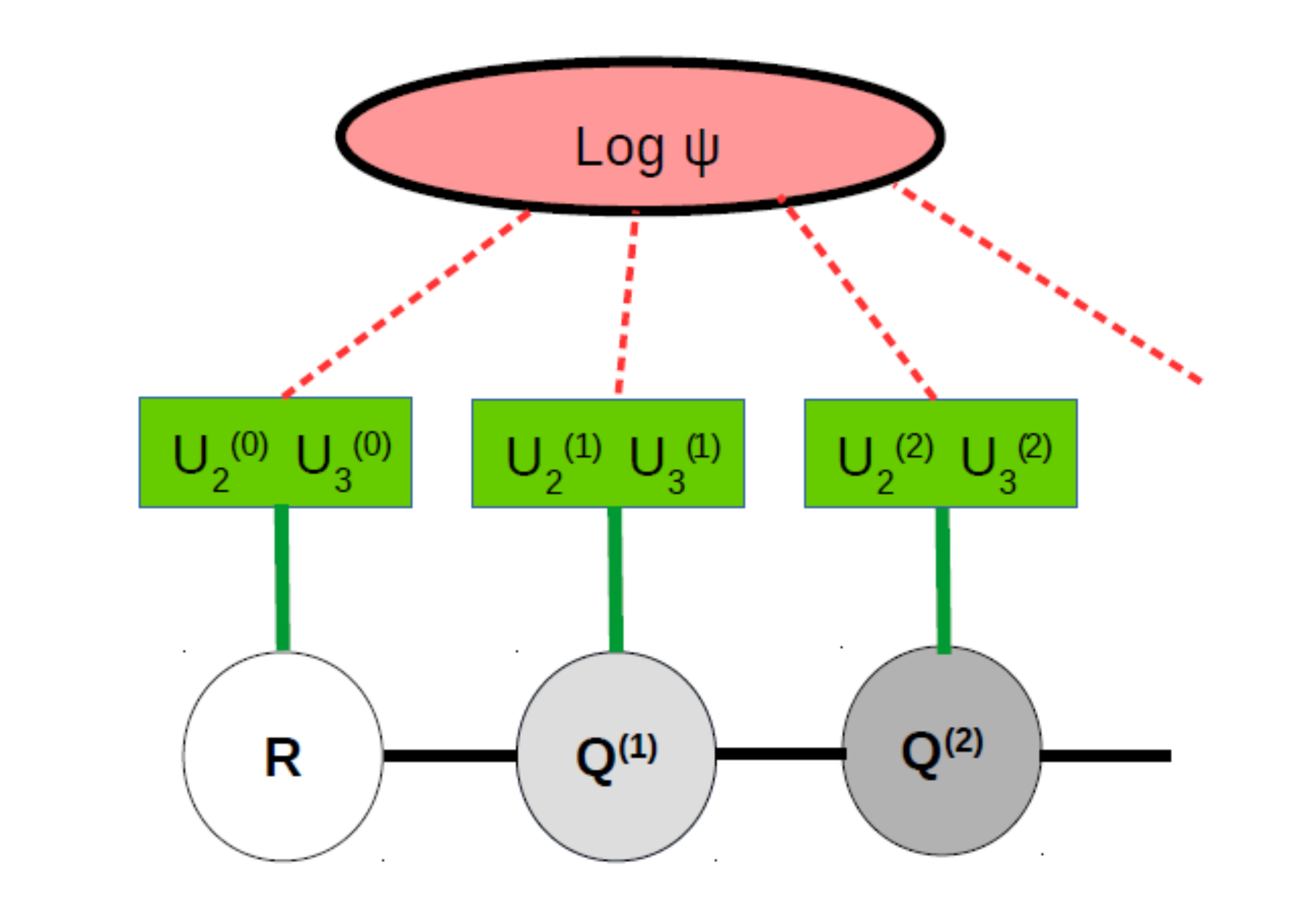}
\caption{Structure of the iterated backflow wave function obtained after approximated integration over the hidden layers of SWF and projector Monte Carlo wave functions (see Fig.\ref{SWF}). Each layer introduces a new set of non-linear functions $U^{(n)}$ (here, two and three-body Jastrow forms, 
$U^{(n)}_2$, $U^{(n)}_3$) and backflow coordinates $\Qvec^{(n)}$ which 
depend only on the coordinates of the previous layers $\Qvec^{(m<n)}$. }
\label{non-lin-net}
\end{figure}

The approximate integration of the hidden layer structure of SWF and projector Monte Carlo wave function can again be considered as a non-linear network, represented in Fig.~\ref{non-lin-net}.
It generalizes the iterative backflow wavefunction employed previously for the description of
two-dimensional fermionic $^3$He \cite{2d} to include also bosonic systems.

Based on hydrodynamic considerations, 
backflow has been introduced originally into wave functions to improve the excitation spectrum
of superfluid $^4$He \cite{bf0,bf1}, but its importance has soon been
recognized for fermionic systems \cite{bf2} where backflow wave functions reduce the fixed-node error in a broad class
of systems. Our heuristic derivation above suggests that the network based on iterative backflow
transformations should rather be considered as a generic description for quantum systems in 
continuous space.
In the following, we explicitly demonstrate that this approach produces high-quality wave functions.

In order to benchmark the performance of the network, we focus
on a system of $N$ $^4$He atoms in a cubic simulation box with periodic boundary conditions. The Hamiltonian for this system is given by
\begin{equation}
H = \sum_i \frac{p^2_i}{2m} + \sum_{i<j} v(r_{ij}).
\end{equation}
and we have use
the HFDHE2 effective potential \cite{pot} for the interatomic interaction, $v(r)$.

\begin{figure}
\includegraphics[width=\columnwidth]{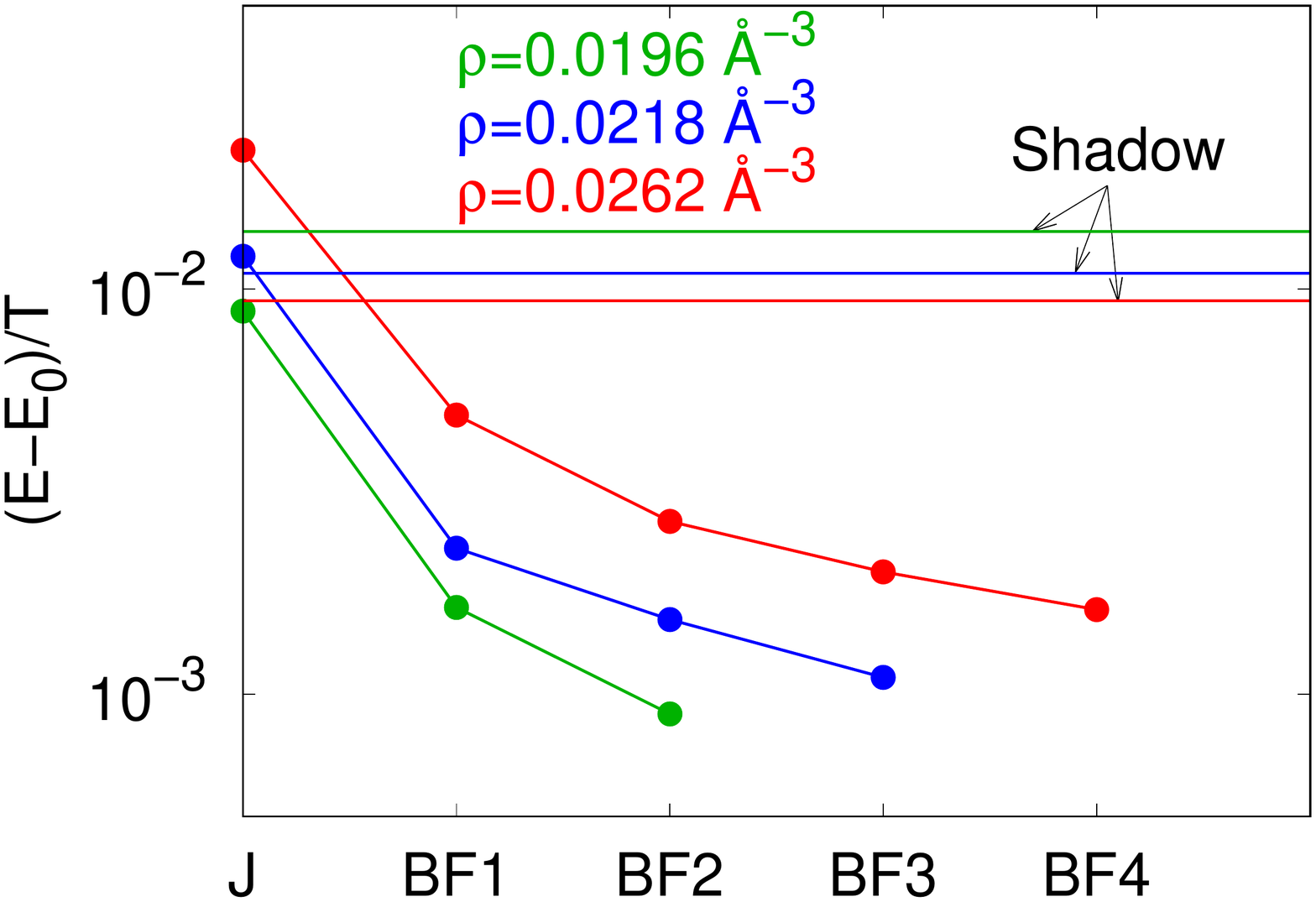}
\caption{Difference between the variational energy and the exact value in units of the kinetic energy $T$ for increasing number of hidden layers $M$ of our non-linear network function for liquid $^4$He in three dimensions.
Simulations are perfomed for $N=64$ atoms at equilibrium density ($\rho=0.0218$ \AA$^{-3}$),
close to freezing ($\rho=0.0262$ \AA$^{-3}$), and for negative pressure ($\rho=0.0196$ \AA$^{-3}$).
Starting from a
Jastrow wave function (J) with two- and three-particle correlations, $M=0$,  the error in the
energy is systematically reduced  increasing the number 
of layers via iterative backflow procedure.}
\label{He4-3D}
\end{figure}

In the network used to describe the ground state of
bosonic $^4$He, each layer $n$ contains two- and three-body correlations in the generalized
Jastrow form
\begin{equation}
\Phi^{(n)}(\Xvec) = e^{-\left(U_2^{(n)}(\Xvec)+U_3^{(n)}(\Xvec)\right)}
\end{equation}
with
\begin{equation}\begin{split}
&U_2^{(n)}(\Xvec) = \sum_{i<j}u_2^{(n)}(x_{ij})\\
&U_3^{(n)}(\Xvec) = \sum_i {\bf G}^{(n)}_i(\Xvec)\cdot{\bf G}^{(n)}_i(\Xvec)\\
&{\bf G}_i^{(n)}(\Xvec) = \sum_j \left( \xvec_i-\xvec_j\right) \zeta^{(n)}(x_{ij})
\end{split}\end{equation}
characterized by one-dimensional functions,
$u_2^{(n)}(x)$ and $\zeta^{(n)}(x)$.
Here, the coordinates $\Xvec$ can refer to either
the bare atomic coordinates ($\Rvec \equiv \Qvec^{(0)}$) or 
the transformed ones ($\Qvec^{(n)}$, $n\ge 1$) which are obtained via
\begin{equation}
\qvec^{(n)}_i = \qvec_i^{(n-1)} + \sum_j \left(\qvec^{(n-1)}_i-\qvec^{(n-1)}_j\right)\eta^{(n)}\left(q_{ij}^{(n-1)}\right)
\label{bf}
\end{equation}
The representations of the one-dimensional functions, $u_2^{(n)}$, $\zeta^{(n)}$, and $\eta^{(n)}$,
establish the network parameters which are determined by energy minimization using
the stochastic reconfiguration method \cite{rocca}.

Although each hidden layer increases the number of variational parameters, 
the scaling of the computational effort for evaluation of the wave function with respect to
the number of atoms, $N$, does not increase \cite{2d}.

In Fig.~\ref{He4-3D} and table \ref{t_en}, we show the error in the ground state energy obtained for $N=64$ $^4$He atoms
in three dimensions. We have considered the liquid at three different densities,
$\rho = 0.0196$ \angm{3} (negative pressure), $\rho = 0.0218$ \angm{3} (equilibrium) and $\rho = 0.0262$ \angm{3} (freezing). The error of the Jastrow or
Shadow wave functions ranges in the tenths of K.
The first backflow layer already results in significantly better variational energies, and additional layers of backflow transformations bring the error down to a few hundredth K.

\begin{table}
 \begin{tabular}{|c|c|c|c|c|}
 \hline
 & \multicolumn{4}{c|}{$\rho = 0.0196$ \angm{3}}\\
 \hline
 & $E_{VMC}/N$ & $\sigma^2/N$ & $E_{SWF}/N$ & $E_{DMC}/N$\\
 \hline
J          & -6.8593(10) & 14.80 &           &            \\
BF1 & -6.9936(14) &  3.03 & -6.765(8) & -7.0243(6) \\
BF2 & -7.0076(15) &  2.14 &           &            \\
%\hline
%Ext.       & \multicolumn{4}{c|}{ -7.033(2)}               \\
Extrap.& -7.033(2)   &       &           &               \\
\hline
 \hline
 & \multicolumn{4}{c|}{$\rho = 0.0218$ \angm{3}}\\
 \hline
 & $E_{VMC}/N$ & $\sigma^2/N$ & $E_{SWF}/N$ & $E_{DMC}/N$\\
 \hline
J          & -6.9137(10)  & 21.40 &           &            \\
BF1 & -7.1204(12)  &  5.22 &           &            \\
BF2 & -7.1367(10)  &  3.30 & -6.937(6) & -7.1691(12) \\
BF3 & -7.1458(14)  &  2.36 &           &            \\
%\hline
%Ext.       & \multicolumn{4}{c|}{ -7.169(3)}                \\
Extrap.& -7.169(3)    &       &           &            \\
 \hline
 \hline
 & \multicolumn{4}{c|}{$\rho = 0.0262$ \angm{3}}\\
 \hline
 & $E_{VMC}/N$ & $\sigma^2/N$ & $E_{SWF}/N$ & $E_{DMC}/N$\\
 \hline
J          & -6.0220(20) & 49.99 &           &             \\
BF1 & -6.4656(25) & 11.20 &           &             \\
BF2 & -6.5230(17) &  9.34 & -6.350(6) & -6.5921(20) \\
BF3 & -6.5402(13) &  5.84 &           &             \\
BF4 & -6.5502(14) &  6.87 &           &             \\
% \hline
% Ext.       & \multicolumn{4}{c|}{ -6.615(2)}               \\
Extrap.& -6.615(2)   &       &           &               \\
 \hline
 \end{tabular}
\caption{
Ground-state energy per particle, in K, of liquid $^4$He in three dimensions at different densities,
obtained with VMC ($E_{VMC}/N$,$E_{SWF}/N$) and DMC ($E_{DMC}/N$)
using different types of trial wave functions: Jastrow wave function without backflow ($J$),
with $n$ iterated backflow transformations (BF$n$) and Shadow Wave Function\cite{SWF}.
We also report the variance $\sigma^2$ of $E_{VMC}$ and the extrapolation of $E_{VMC}/N$ to 
zero variance\cite{2d}. Statistical uncertainties on the last digit(s) are given in parentheses. Tail corrections are calculated assuming $g(r)=1$ for distances larger than half the side of the simulation cell.
}
\label{t_en}
\end{table}

Apart from the energy, also the variance of the local energy, $\sigma^2$, can be computed
at each iteration level, $M$, without additional
computational costs. 
Under suitable conditions \cite{2d} the extrapolation of the energy to $\sigma^2=0$ with a leading linear term gives the
exact ground state energy. 
The largest error in the extrapolated values is -0.02K at the highest density.

\begin{figure}
\includegraphics[width=\columnwidth]{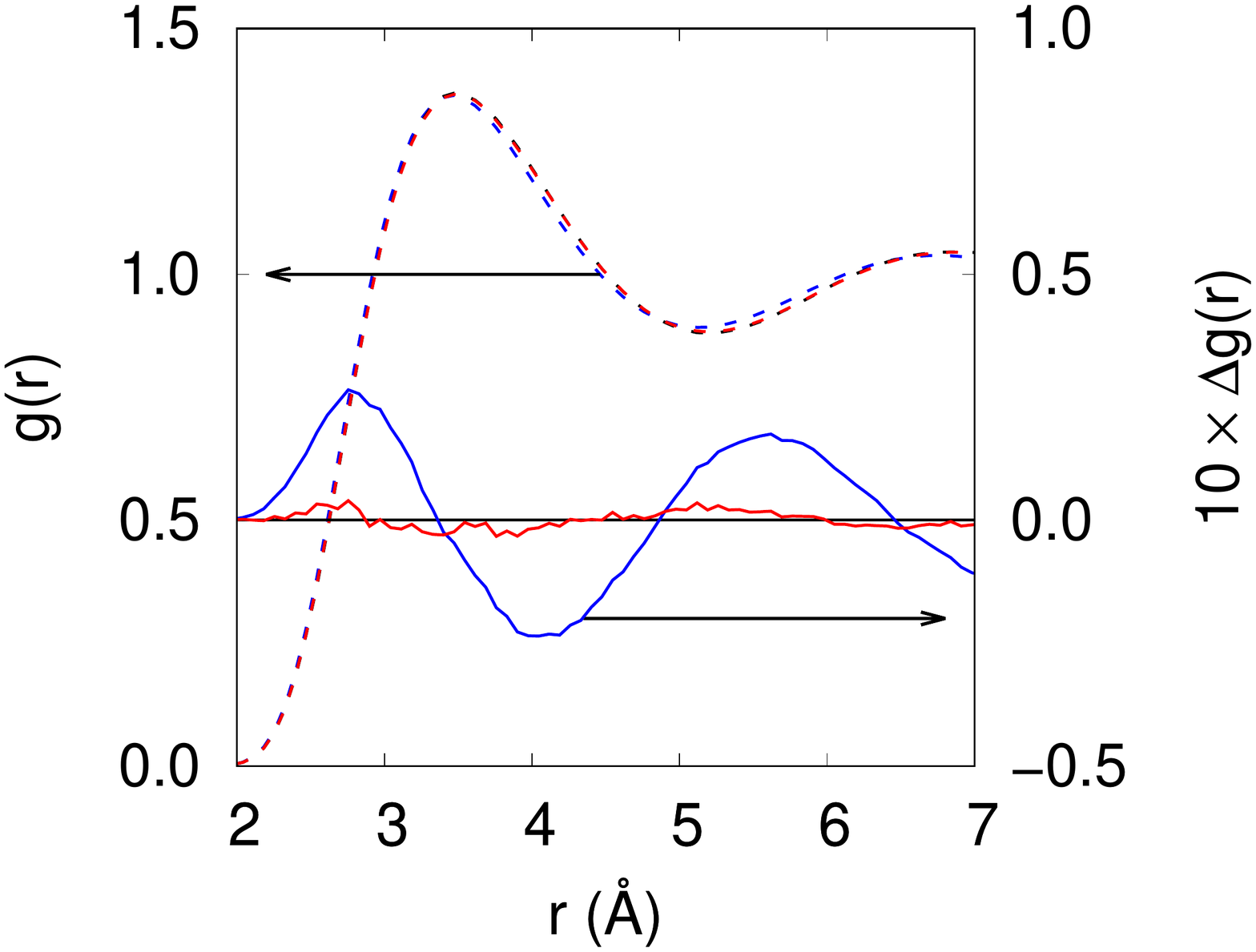}
\caption{
Pair correlation functions $g(r)$ for liquid $^4$He in three dimensions at equilibrium density. Dashed lines (left scale) show variational results without 
backflow terms (blue) and with three backflow iterations (red), as well 
as DMC results (black, barely visible behind the red dashes; extrapolated
estimate\cite{mitas} using the BF3 trial wave function).
Solid lines (right scale) show a tenfold magnification ot the deviation
between the VMC and DMC results.
}
\label{gr}
\end{figure}

A roughly tenfold increase in accuracy is also obtained in the
pair correlation function $g(r)$, as shown in Fig.~\ref{gr}
for the equilibrium density. 

\begin{figure}
\includegraphics[width=\columnwidth]{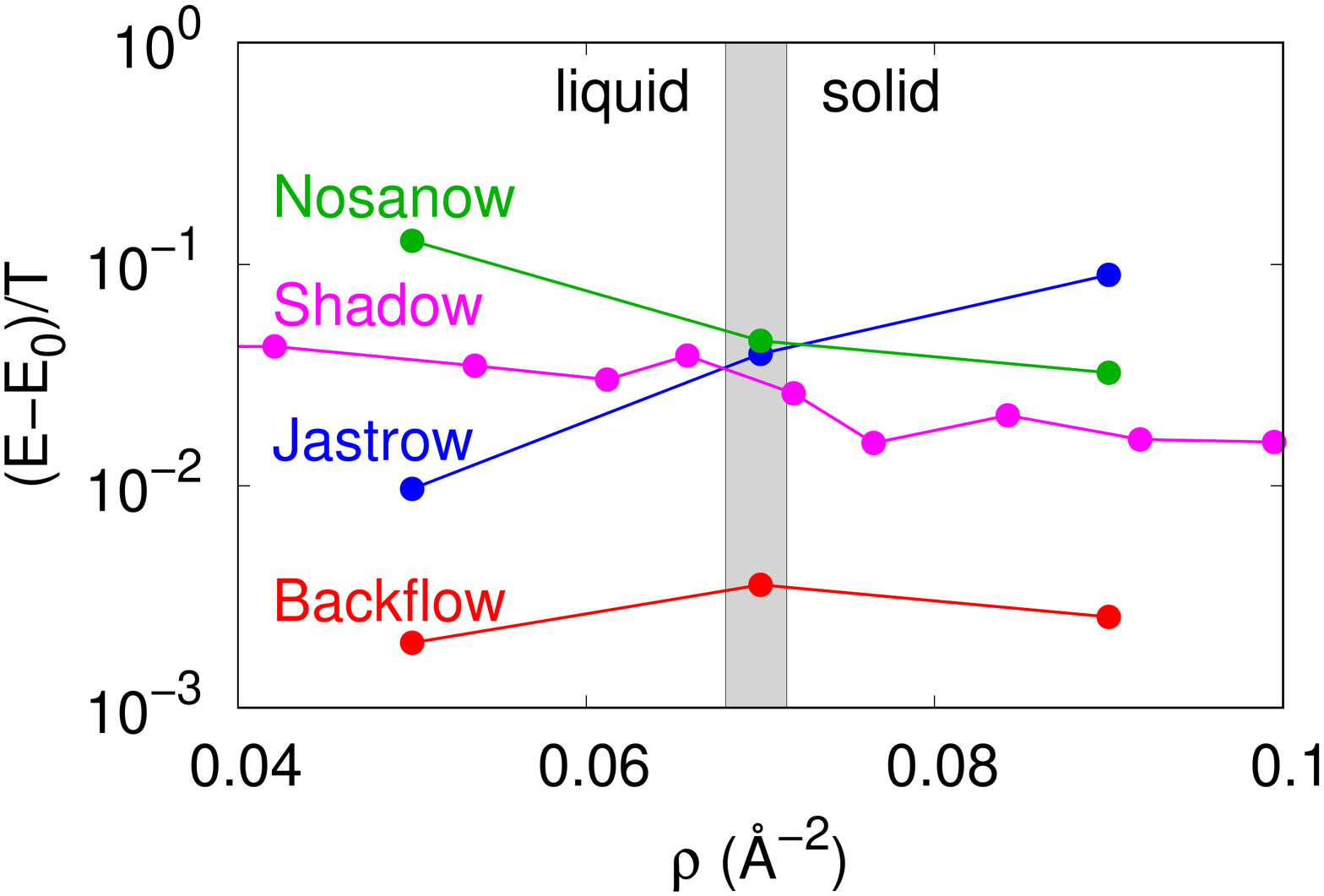}
\caption{Error of the ground state energy of liquid and solid $^4$He in two  dimension in units of
the kinetic energy $T$ across the
the liquid-solid coexistence region
(shaded)\cite{gordillo}, using various wave functions: Shadow\cite{chester}, Jastrow, Nosanow and iterative backflow network (5 iteration layers at coexistence, 4 layers otherwise).}
\label{He4-2D}
\end{figure}

In order to describe freezing, the liquid and the solid phase must usually be
described by different functional forms within VMC, as the usual Jastrow wave function is in general unable to localize the atoms in a crystal (unless the pair pseudopotential is made unreasonably hard). This bias propagates even to projector Monte Carlo methods (DMC) based on 
importance sampling using the Jastrow function. In order to correctly describe the solid phase,
 usually one uses in VMC and DMC an unsymmetrized Nosanow wave function \cite{nosanow} where the atoms are individually tied to predetermined lattice sites by a one-body term. In this setup, the Jastrow(Nosanow) phase describes a metastable liquid(solid) phase at densities higher(lower) than the coexistence region. 

One important conceptual progress of SWF was the possibility to describe both liquid and solid
$^4$He within the same wave function \cite{vitiello}, without breaking translational invariance or Bose symmetry. Remarkably, this feature is shared by our network wave function. In Fig.~\ref{He4-2D}, we compare the performance of 
network, Jastrow-Nosanow, and Shadow wave functions for $N=16$ $^4$He atoms in two dimensions around the liquid-solid transition\cite{gordillo}.
Again, our backflow network function achieves a roughly tenfold reduction of the variational error with respect to  Shadow\cite{chester}, Jastrow, and Nosanow wave functions --over a large density range and across a phase transition. 
For the higher density, $\rho = 0.09$, we make sure that the backflow wave function describes a solid by inspection of the pair distribution function
 $g(x, y)$ calculated with VMC, which is hardly distinguishable from the Nosanow
(bona fide solid) result. For the lowest density $g(x, y)$ turns into a radial, liquid-like pair distribution function, while at coexistence it is intermediate between the Nosanow and Jastrow results, much closer to the former.

\begin{table}
 \begin{tabular}{|c|c|c|c|c|}
 \hline
 & \multicolumn{4}{c|}{$\rho = 0.01635$ \angm{3}}\\
 \hline
 & $E_{VMC}/N$ & $\sigma^2/N$ & $E_{FNDMC}/N$ & $E_{EXP}/N$ \\
 \hline
J   & -1.6812(17) & 38.23 & -2.0925(16) &            \\
BF1 & -2.0844(13) & 16.70 & -2.2760(10) &            \\
BF2 & -2.2278(23) &  8.10 & -2.3190(14) & -2.481     \\
BF3 & -2.2576(15) &  5.60 & -2.3288(14) &            \\
%\hline
%Ext.       & \multicolumn{4}{c|}{ -2.35(1)}               \\
Extrap.& -2.34(1)    &       &             &               \\
 \hline
Ref.~\cite{manyBodyCorr} & -2.168(3) & 14 & -2.306(4) & \\
\hline
 \hline
 & \multicolumn{4}{c|}{$\rho = 0.02380$ \angm{3}}\\
 \hline
 & $E_{VMC}/N$ & $\sigma^2/N$ & $E_{FNDMC}/N$ & $E_{EXP}/N$\\
 \hline
J   &  0.7582(32)  & 111.06 & -0.2890(43) &            \\
BF1 &  0.0952(17)  &  65.03 & -0.5572(41) &            \\
BF2 & -0.3661(42)  &  29.43 & -0.6545(15) & -0.918     \\
BF3 & -0.5177(32)  &  17.47 & -0.6911(16) &            \\
BF4 & -0.5531(27)  &  14.41 & -0.7013(42) &            \\
%\hline
%Ext.       & \multicolumn{4}{c|}{ -0.746(9)}                \\
Extrap.& -0.723(3)    &        &             &                \\
 \hline
Ref.~\cite{manyBodyCorr} & -0.127(5) & 49 & -0.6485(4) & \\
 \hline
 \end{tabular}
\caption{
Ground-state energy per particle, in K, of liquid $^3$He in three dimensions at equilibrium and freezing densities,
obtained with VMC ($E_{VMC}/N$) and fixed-node DMC ($E_{FNDMC}/N$)
using different types of trial wave functions: Jastrow wave function without backflow ($J$),
and with $n$ iterated backflow transformations (BF$n$).
We also report the variance $\sigma^2$ of $E_{VMC}$, the extrapolation of $E_{VMC}/N$ to 
zero variance\cite{2d}, and the experimental value $E_{EXP}/N$ \cite{exp}. Statistical uncertainties on the last digit(s) are given in parentheses. Tail corrections are calculated assuming $g(r)=1$ for distances larger than half the side of the simulation cell. 
The results from entry S3BF4 of table I of Ref.~\cite{manyBodyCorr} are corrected by a perturbative estimate of the difference due to their use of 
a different \cite{korona} pair potential $v(r)$.
}
\label{t_he3}
\end{table}
Up to now, we have demonstrated the quality of our backflow network to describe bosonic quantum
systems, where stochastic projection Monte Carlo methods provide exact results for benchmarking.
Now, we show that our approach significantly improves the description of strongly correlated fermions
in three dimensions, similar to previous results \cite{2d} obtained for two dimensional liquid $^3$He.
In table \ref{t_he3} we list estimates of the ground state energy obtained with different
wave functions for $N=66$ $^3$He atoms in three dimensions, at equilibrium and freezing density.
The previous best estimates \cite{manyBodyCorr} were obtained introducing explicit correlations 
up to four-particle in the Jastrow factor and 
three-particle in the backflow coordinates. The results from Ref. \cite{manyBodyCorr} included in table  \ref{t_he3}
refer to a spin-singlet pairing wave function, which performs
marginally better than a Slater determinant of plane waves. They lie between BF1 and BF2, 
showing that the implicit inclusion of correlations at all orders through backflow iteration is more effective than
explicit construction of successive $n$-order terms (although nothing prevents the two approaches to be combined). 
Furthermore, systematic improvement is more easily obtained by adding further 
layers of backflow transformations than further explicit correlations.

In the lack of exact benchmark results for this fermionic case, we compare our results
to the experimental equation of state \cite{exp}. The HFDHE2 pair potential adopted here is accurate within a few hundredth of a K
from equilibrium to freezing density for $^4$He \cite{kalos}, and presumably equally reliable also for $^3$He at slightly
lower densities; furthermore the number of particles, $N=66$, is chosen to give a small finite-size shell effect on the kinetic 
energy, so that the energies in the table are reasonably close to the thermodynamic limit.
Our best estimate, the extrapolation to zero variance, is higher than the experimental energy by 0.14 K at equilibrium
density, and by 0.19 K at freezing. This comes to a surprise, as 
for $^4$He in two and three dimension and small systems of $^3$He in two dimensions --very similar 
cases where exact results are available-- the error in the zero-variance extrapolation is of order of 0.01 K.
Nevertheless the improvement over the previous variational and fixed-node energies remains significant.

In summary, in this paper, we have put the iterated backflow description 
%applied previously to two dimensional liquid $^3$He
 \cite{2d} in a more general frame. We have
demonstrated the quality of our backflow network for quantitative description 
of bosonic and fermionic helium systems in two and three dimensional continuous space.

In our description, each backflow transformation corresponds to a hidden layer, and
each new layer depends on the coordinates in the previous one.
This structure is motivated heuristically from a Path-Integral projection.
However, from the interpretation as neural network, one may ask if our description of several
fully connected  layers can be simplified or made more efficient, e.g. replace the multiple layer
structure by a single layer with $M$ times more backflow coordinates or find a formulation
closer to a restricted Boltzmann machine as recently applied to discrete lattice systems \cite{CT,NNHubbardFermi}.

Although properties of the ground state or systems in thermal equilibrium
can be exactly computed by stochastic algorithms for bosonic systems \cite{RMP}, 
accurate explicit expressions
for correlated quantum states open the possibility to study also
out-of-equilibrium time evolution of a many-body system \cite{TDVMC,cTDVMC} 
in two or three dimensional, continuous space.

\end{document}